\documentclass[usegraphicx,useAMS,usenatbib]{mn2e}

\title{The non-ballistic  superluminal motion in the plane of the sky}

\author[Biping Gong]
    {Biping Gong$^{1}$\thanks{E-mail: bpgong@mail.hust.edu.cn}\\
$^1$Department of Physics, Huazhong University of Science and
Technology, Wuhan 430074, China, bpgong@mail.hust.edu.cn \\}

\begin{document}

\maketitle

\date{Accepted. Received}
\voffset-0.4in
\label{firstpage} \maketitle
\pagerange{\pageref{firstpage}--\pageref{lastpage}} \pubyear{2008}
\def\LaTeX{L\kern-.36em\raise.3ex\hbox{a}\kern-.15em
    T\kern-.1667em\lower.7ex\hbox{E}\kern-.125emX}

\begin{abstract}
{Faster-than-light or superluminal motion was originally predicted
as a relativistic illusion of ballistic moving ejecta, and
confirmed in a few tens of sources observationally. However, the
recent results of the long-term multi-epoch observations of
quasars, active galaxies, tracing the structure further along the
jets and following the motion of individual features for longer
time, rise questions that are difficult to understand by the
standard ballistic model. I.e., the ejecta are aligned with the
local jet direction, instead of the core; and within individual
jets apparently inward-moving features are observed. Here we show
that these unexpected phenomena, although only a small fraction
among large samples, indicate the existence of non-ballistic jet
motion, in which a continuous jet produces a discrete  hot spot.
And the precession of such a hot spot in the plane of the sky
appears superluminal. Therefore, an unified and simple
interpretation to the new results is obtained, which can be
further tested through its predictions on the evolution of ejecta.
The study is of importance in the understanding of the nature of
superluminal motion,  the interaction of jets and surrounding
materials, as well as the common physics underlying quasars and
microquasars. }
\end{abstract}


\section{Introduction}

The appearance on the sky of relativistically moving out flow,
which expands at speed greater than the speed of light, has been
predicted by Rees (1966), five years in advance of the discovery
of superluminal motion~\citep{Rees66, Whitney71, Cohen71}. In
units of the speed of light, $c$, the apparent speed $\beta_a=v/c$
of separation is the ratio of the difference in observed positions
to the observed time-interval. With the blob travelling at angle
$\theta$ with respect to the line of sight (LOS), $\beta_a$ is
given by
\begin{equation}
\label{vapp2}
\beta_{a}=\frac{\beta_p\sin\theta}{1-\beta_p\cos\theta}
 \,,
\end{equation}
When $v\sim c $, and $\theta$ at certain range, the apparent
speed, $\beta_{a}$, can be larger than the speed of light. In fact
Eq.~($\ref{vapp2}$) and Fig.~1a correspond to a simple ballistic
motion of ejecta, with a fixed jet axis, which has been applied to
the superluminal motion of quasar like 3C279.

To interpret the S-shape pattern of SS433, a modified ballistic
model has been proposed~\cite{Abell79}, in which the direction of
the jet axis is variable, and denoted by $a$, $b$, and $c$ in
Fig.~1b. In this model, features $b$ and $b^{\prime}$ in Fig.~1b,
are the same ejection evolving at different times, which should
always move along the jet axis  $o-b$. As the jet axis precesses
to another direction, i.e., $o-c$, another feature ejects out.
Similarly, the trajectory of it should always be fixed at that
direction.
Assuming that only one feature is
ejected at each direction, then the connection of these features
ejected at different directions like $a$, $b$ and $c$ forms a one
sided spiral pattern, as in Fig.~1b. And considering the receding
features at opposite direction to $b$ and $c$  a S-shape pattern,
like SS433, is explained.

Generally, the ballistic model, with jet axis fixed or not,
implies that each individual feature must move along the jet axis
away from the core. Thus, the observed inward motion of features
can be interpreted by newly emerging jet features ejecting from
the core at the right position and time, which mimics the decrease
of the apparent separation of the core and other features.

\begin{figure}
\includegraphics[width=0.45\textwidth]
{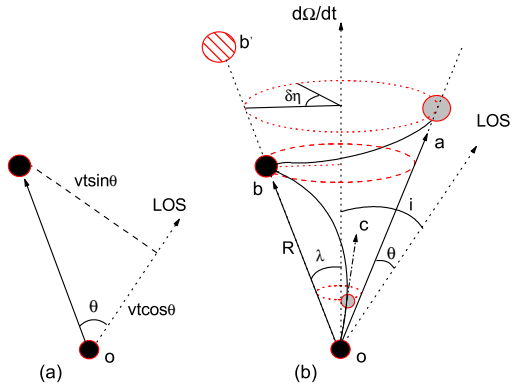}
\caption{\small A schematic illustrating of three models of
superliminal motion. The first one is the pure ballistic model
with a fixed jet axis, shown at the left side. The second one is
the model of ballistic plus precession, represented by the solid
spiral connecting the features, a, b, and c. And the last one is
the non-ballistic model, denoted by the dashed ellipse through
feature b, which precesses with a constant distance, $R_d$, to the
core. \label{fig1} }
\end{figure}

Therefore, the ballistic model implies that  the whole curved
pattern would move further away from the core at the next step of
evolution due to that each individual feature evolves linearly as
in Eq.~($\ref{vapp2}$). However, this is not consistent with the
observation that  the jet flow apparently occurs along preexisting
bent channels~\cite{Kellermann04}, instead of away from the core
as a whole. Therefore, it is conceivable to consider other
possibilities  of jet motion.

\section{The non-ballistic model} \label{sec:pre+lc}
Theoretically,  a hot spot  can be generated by the interaction of
a continuous jet with the surrounding material.
The process of  continuous ejection producing ``discrete hot
spots" can be simply extracted from Dermer (1999).
The energy of the injected nonthermal particles comes at the
expense of the directed bulk kinetic energy of the fluid. Assuming
that the system produces an outflow with total energy $E_0$ and
initial bulk Lorentz factor $\Gamma_0$. Since most of the energy
of the flow is bound up in the kinetic energy of baryons, assumed
to be protons, then $E_0=\Gamma_0 N_{t}m_pc^2$, with $N_{t}$ the
total number of protons~\citep{Dermer99}.

Through the dynamics of the blast wave, which decelerates by
sweeping up material from the surrounding medium, the deceleration
radius, which can be considered as the distance of a hot spot to
the core, is obtained as~\citep{Dermer99},
\begin{equation}
\label{Rd} R_d=[\frac{(3-\zeta)E_0}{4\pi f_b
n_0\Gamma_0^2m_pc^2}]^{1/3},
\end{equation}
where  the density of external medium can be parameterized by the
expression $n_{ext}(R)=n_0(R/R_d)^{-\zeta}$. In the simplest case,
one finds that $\Gamma(R)\cong\Gamma_0$ for $R\ll R_d$, and
$\Gamma(R)\propto R^{-g}$ for $R\gg R_d$, where $g=3-\zeta$ and
$g=(3-\zeta)/2$ in the adiabatic and radiative regimes,
respectively. The term $f_b$ represents the fraction of the full
sky into which the explosion energy is ejected. The deceleration
radius, $R_d$, occurs at the observing time~\citep{Dermer99},
\begin{equation}
\label{time}
t_d=\Gamma_0(1-\beta_0\cos\theta)(1+z)R_d/(c\Gamma_0),
\end{equation}
where $\beta_0=\sqrt{1-\Gamma_0^{-2}}$ and $z$ is the redshift and
$\theta$ is the misalignment angle between the jet and LOS. At a
given  frequency and  $\theta$, there is a peak in flux density at
the time $t_d$ of Eq.~($\ref{time}$), in other words, the
brightest emission region during its whole evolution time appears
at $t_d$ and distance $R_d$ to the core.



The light curves for the model synchrotron flare at X, gamma, and
radio frequencies, both along the jet axis (thick curves), and at
$20$deg to the jet axis (thin curves) are shown in Fig.~2 of
Dermer (1999). Obviously, each peak in the light-curve corresponds
to a hot spot during its whole evolution time.


The radial size of such a hot spot can be defined as: the length
corresponding to the decline of the flux density from the peak for
one order of magnitude (for X-ray emission), or for a factor of
five below the peak (for radio emission). Thus, the radial length
of a hot spot compared with its distance to the core is  $\delta
R_d/R_d=\delta t/t$, which is between $10^{-1}$ to $10^{-3}$ in
Fig.~2 of Dermer (1999).

When the jet axis precesses to another direction, the process of
ejection and deceleration repeats. Hence another hot spot is
produced, and its distance to the core is still $R$ in the case of
an isotropic distribution of the material. Thus different hot
spots are generated by the same continuous jet, interacting with
matter at different directions, which mimics one hot spot moving
in space continuously.

Contrarily, for the ballistic model hot spots or features  are
produced by discrete ejection events. The distance of each feature
to the core is determined by Eq.~($\ref{vapp2}$), which varies
with time.
Thus simply replacing the radial velocity of a feature, ${\bf v}$,
of the ballistic model of SS433~\cite{Abell79,Hjellming81}, by the
distance of it to the core, $R_d$, the  precession of a hot spot
under the non-ballistic model is obtained as,
\begin{eqnarray}\label{lk}
 R_{x} &=& R_d[\sin\lambda\sin i\cos\eta +\cos\lambda\cos i ] \,, \nonumber\\
 R_{y} &=& R_d[\sin\lambda\sin\eta ]\,, \nonumber\\
 R_{z} &=&R_d[\cos\lambda\sin i -\sin\lambda\cos i\cos\eta ]
\,,
\end{eqnarray}
where $R_{x}$, $R_{y}$, and $R_{z}$ are the components of $R_d$ in
the coordinate system $x-y-z$. The $x$-axis  is towards the
observer, rotating around the $x$-axis for angle $\xi$, so that
the new $y$-axis ($\Delta\delta$) will point north, and the new
$z$-axis ($\Delta\alpha$) will point east. As shown in Fig.~1,
$\lambda$ is the opening angle of the precession cone, $i$ is the
inclination angle between the jet rotation axis and  LOS, and the
precession phase is $\eta=\dot{\Omega} t+\eta_0$ ($\eta_0$ is the
initial phase). The displacement of a feature to the core can be
described  with respect to $\Delta\delta$ and $\Delta\alpha$ as,
\begin{eqnarray}\label{lalpha}
R_{\alpha} &=& R_y\sin\xi+R_z\cos\xi\,, \nonumber \\
R_{\delta} &=& R_y\cos\xi-R_z\sin\xi \,.
\end{eqnarray}
Differentiating Eq.~($\ref{lalpha}$), the  velocity projected to
the plane of the sky is given by,
\begin{eqnarray}\label{1915v}
v_{\alpha} &=& \dot{R}_y\sin\xi+\dot{R}_z\cos\xi\,, \nonumber \\
v_{\delta} &=& \dot{R}_y\cos\xi-\dot{R}_z\sin\xi \,.
\end{eqnarray}
The time taken by a spot to precess for tangent distance of the
size of a hot spot, $\delta R_d$, is $\delta t=\delta
\eta/\dot{\Omega}$, where $\delta\eta=\delta R_d/(R_d\sin
\lambda)$. The cooling time of a hot spot, $\delta t_{co}$,
corresponding to, i.e., the time taken for the radio peak to
decline for a factor of five, can be inferred from Fig.~2 of
Dermer (1999). If $\delta t<\delta t_{co}$, then the hot spot
appears as a filament, otherwise, it appears as a hot spot.
Obviously the precession can make a filament bent or twisted.

\section{The apparent luminosity } \label{sec:lum}
With the intrinsic luminosity is $L_0$, the apparent luminosity of
a jet feature is given by,
\begin{equation}\label{L}
L=L_0\delta^n  ,
\end{equation}
where $\delta=\gamma_b^{-1}(1-\beta_b\cos\theta)^{-1}$, and
$\gamma_b$ is the Lorentz factor, given by
$\gamma_b=(1-\beta_b^2)^{-1/2}$. And $n$ depends on the geometry
and spectral index, which is typically in the range between 2 and
3. The angle $\theta$, denoting the misalignment angle between the
LOS and the jet axis, changes as the  precession of the jet axis,
\begin{equation}\label{theta}
 \cos\theta = \cos\lambda\cos i +\sin\lambda\sin i \cos\eta  \,.
\end{equation}
As  $\theta$ varies, a feature appears moving away from the core
or close to the core at different precession phase,
as shown in Fig.~1. Hence, the apparent luminosity varies with the
precession. Obviously, the maximum luminosity occurs when
$\theta\sim 0$.


Whether a feature  is detectable or not is dependent of  the
misalignment angle between the jet axis (corresponding to the hot
spot) and the LOS, $\theta$, as given  by Eq.~($\ref{theta}$). The
larger the angle, $\theta$, the weaker the Doppler boosting
effect. The jet emission becomes undetectable when  $|\theta|$
exceeds certain limit. This explains the disappearance of features
at large angular displacement, i.e., for GRS
1915+105~\cite{Miller-Jones07}.


By the new model the apparent velocity of a hot spot, $\beta_a$,
can be represented as,
\begin{equation} \label{betaa}
\beta_a\propto\frac{\dot{\Omega}R\sin\lambda}{c}
 \,,
\end{equation}
which is dependent of parameters such as: $R$, $\dot{\Omega}$,
$\lambda$, $i$, and $\eta$. Thus, the corresponding pattern speed,
$\gamma_p$, if inferred from Eq.~($\ref{vapp2}$), is not directly
related with the bulk speed, $\gamma_b$. This explains the
observation of both the slow quasars with $\gamma_p\ll\gamma_b$
and the fast quasars with $\gamma_p\sim\gamma_b$~\cite{Cohen07}.

\section{Inward motion and bent trajectory} \label{sec:inward}

The apparent inward motion for quasars~\cite{Kellermann04}
can be explained in the context of non-ballistic model. For some
samples, the LOS is as the left side of Fig.~2, denoted by
$LOS_1$, in which ejecta are moving out of the core ($A_2$ and
$B_2$ are projected to 1 and 2, respectively); for other ones, the
observer's view is as $LOS_2$, at the right side of Fig.~2, where
the ejecta are moving towards the core ($B_2$ and $C_2$ are
projected to 3 and 4, respectively). Under such a circumstance,
spot $C_2$ would precess to $D$, which appears moving toward to
the core, by projecting to $D^{\prime}$ on the plane of sky.

\begin{figure}
\includegraphics[width=0.45\textwidth]{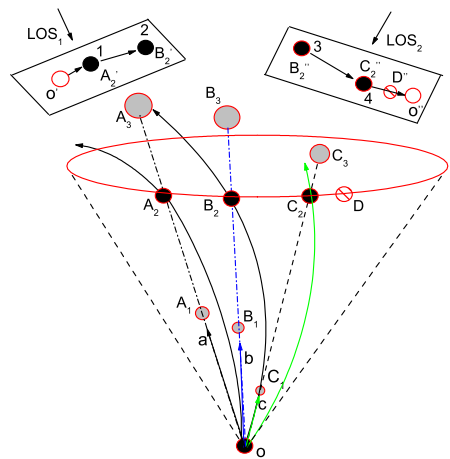}
\caption{\small A schematic illustrating of inward motion under
the non-ballistic model. The precession of jet axis results in
motion of  the brightest feature in the elliptic trajectory. For
some sources, the observer's view angle is as  the left side which
appears that ejecta are moving out of the core ($A_2$ and $B_2$
are projected to 1 and 2 respectively); whereas for other sources,
it is as the right side in which ejecta seem to move inwardly to
core ($B_2$ and $C_2$ are projected to 3 and 4 respectively). The
next evolution would be that $C_2$ moves to D which is projected
to $D^{\prime}$ which is even closer to the core. }
 \label{Fig:fig2}
\end{figure}

Beside a qualitative interpretation of apparently inward motion,
as in Fig.~3, the observation of quasar 0119+041 can also be
fitted by the new model. Assuming the parameters
$\dot{\Omega}=43.0$deg/yr, $R_d=2.9$mas, $i=11.0$deg,
$\lambda=3.7$deg, $\eta=225.0$deg, and $\xi=4.3$deg, and putting
them into  Eq.~($\ref{lalpha}$), Fig.~3 can be obtained.
Obviously, the curve given by the new model fits the observations
(dots) better than the dashed line of the ballistic model, shown
in Fig.~3. The precession velocity of the jet axis,
$\dot{\Omega}$, can originate from either the Lense-Thirring
effect, or the torque exerted by the companion object in a binary
system~\cite{Katz97}.

\begin{figure}
\includegraphics[angle=-90,width=0.45\textwidth]{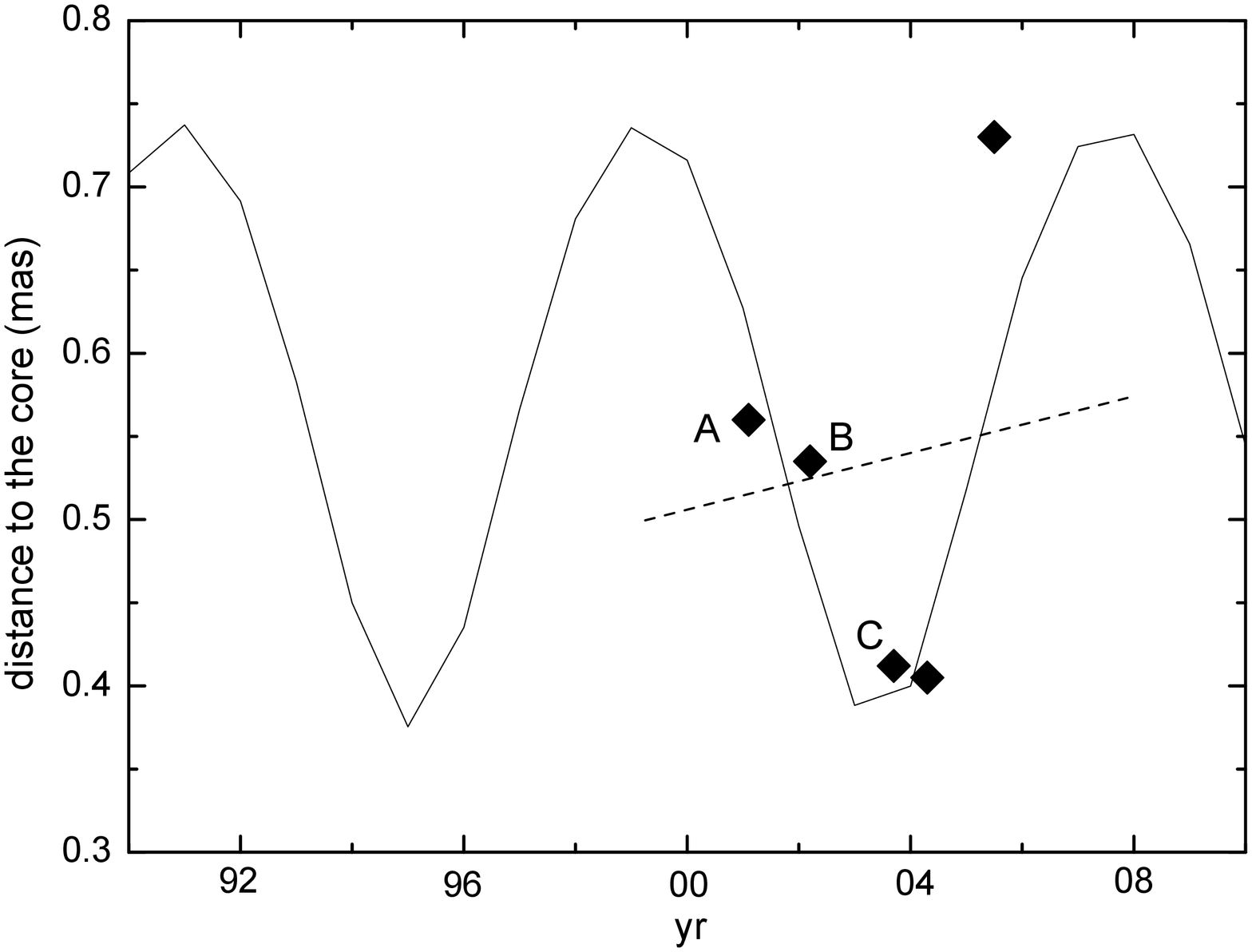}
\caption{\small Fitting of 0119+041. The observational
inward-moving features of 0119+041 are represented by dots, which
are fitted by the curve obtained by putting the parameters of
0119+041 in the text into Eq.~($\ref{lalpha}$). The dashed line is
the fitting by the ballistic model. \label{Fig:fig3} }
\end{figure}

The observed bent trajectories~\cite{Kellermann04} can also be
explained by the new model. According to Eq.~($\ref{lalpha}$), the
whole trajectory of a hot spot on the plane of the sky in one
precession period is an ellipse. And part of this elliptic
trajectory appears as a bent trajectory, which should be nonlinear
and non-radial. This explains why the ejecta are aligned with the
local jet direction (as 1-2, or 3-4), instead of the core ($o$-2,
or 3-$o$), as shown in Fig.~2.

%

\section{Multi-ejection} \label{sec:multi}

The creation of multi-components in pulse profiles of a pulsar may
result in a nested cone structure~\cite{Gil93}, or a patchy beam
structure~\cite{Lyne88}. If the continuous jet producing hot spots
in quasars and microquasars has  a similar clumps beam, then the
interaction of such a beam (or structured jet) with the
interstellar material can reproduce multi-components either. In
such case, each component can have its own misalignment angle with
the axis of the precession cone, and its distance to the core can
also be different.

It is conceivable that each component has approximately the same
precession velocity  around the axis of precession cone. So that
the precession period of the major beam around the axis of
precession cone and the pattern of the major beam are unchanged
during jet precession. The precession of such a jet mimics a
multi-component (or sub-spot) moving in space as in Fig.~4.

\begin{figure}
\includegraphics
[width=0.42\textwidth]
{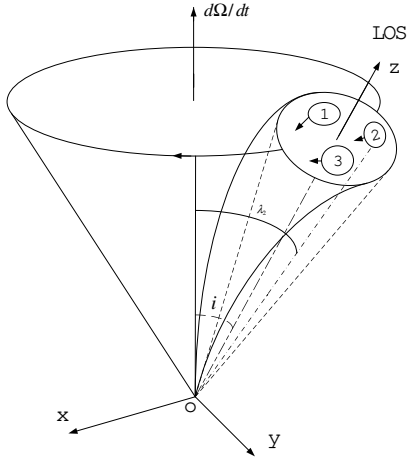} \caption{\small A schematic illustrating of multi-sub
hot spot scenario. Each one has its own misalignment angle with
respect to the jet axis, i.e., $\lambda_2$ for sub-hot spot 2, and
its distance to the core, $R_k$, can also be different. In this
figure, sub-hot spot 1 and 3 appear to move away from the core,
denoted by o, whereas, spot 2 appears to move inwardly to the
core. \label{fig4} }
\end{figure}
By assuming the parameters of Table~1, the multi-ejection of
microquasar GRS 1915+105, can  be fitted as in Fig.~5.
\begin{figure}
\includegraphics[width=0.45\textwidth]{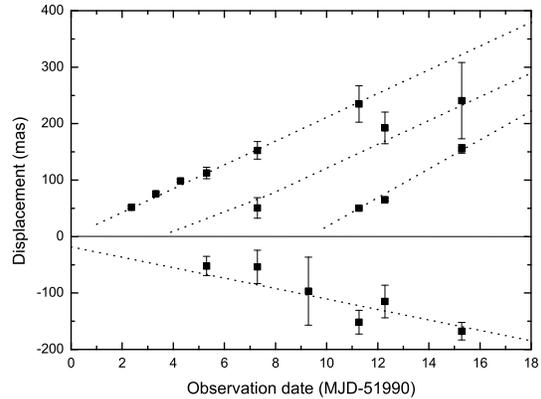}
\caption{\small The multi-ejection of GRS 1915+105 can be fitted
by the parameters of GRS 1915+105 in Table~1, in which the three
sub-hot spots satisfy $\lambda\approx i$,  so that they appear as
ejected from the core.
 \label{Fig:fig5}}
\end{figure}
As the beam precesses through LOS, only these sub-hot spots, with
the opening angle of precession cone, $\lambda_k$, equal the
inclination angle, $i$, correspond to a zero angular displacement
to the core, which appear to be ejected from the core.  Those with
$\lambda_k\neq i$, precess through LOS with nonzero angular
displacement and hence don't appear to be ejected from the core,
which explains the phenomena i.e., on 0736+017, 0735+178 and
1219+285~\cite{Kellermann04}.

\begin{table}
\begin{center}
\caption{The fitting parameters of and GRS 1915+105. }
\begin{tabular}{cccccc}
\hline \hline
  $\dot{\Omega}$ & $i$ & $\xi$ & $\lambda_1$
  &  $\lambda_2$
&  $\lambda_3$ \\
0.283 & 52.9 & 285.0 & 52.9 & 52.9 & 52.9 \\
$\eta_1$ & $\eta_2$ & $\eta_3$ &
 $R_1$ &  $R_2$ &  $R_3$ \\
0.0 & -1.2 & -2.6 & 5350.0 & 5350.0 & 6500.0 \\
\hline
\end{tabular}
\end{center}
{\small $\dot{\Omega}$ is in unit  deg/day. $R_1$, $R_2$, and
$R_3$ are the distances of the three sub-hot spots to the core in
unit of mas. All other parameters are in unit of degree. The
receding feature of GRS 1915+105 is obtained by assuming
$R_r=2675.0$mas, and $\eta_r=5.5$deg. }
\end{table}

\section{Precession of a bent jet} \label{sec:bentjet}

The source 0735+178 is a bizarre object, since  all strange
phenomena occur on it: inward motion (square dots), not from the
core (all dots) and ejecta aligning with the local jet direction,
instead of the core (all dots), as shown in Fig.~6. Moreover, it
contains both outward motion (dots in filled and hollow circles)
and inward motion (square dots).

Under the new model, the jet motion of 0735+178 results in the
precession of a bent jet, which has three hot spots instead of one
in 0119+041. The three hot spots can be seen as $C_1$, $B_2$, and
$A_3$ in Fig.~2, locating at different radial distances, $R_1$,
$R_2$, and $R_3$ respectively. The hot spots corresponding to
$R_1$ and $R_2$ are not bent much, when the bulk velocities
(radial), $v_1/c$ and $v_2/c$ are much larger than the precession
induced tangent velocities, $\beta_1$ and $\beta_2$, given by
Eq.~($\ref{betaa}$). As the distance of the plasmoid to the core
becomes greater than $R_2$, its velocity drops considerably, so
that $v_3/c$ is less (or much less) than that of precession
induced velocity, say, $\beta_3$. Thus the hot spot $A_3$, appears
bent significantly. The image of 0548+165 may provide an example
of such significantly bent jet~\cite{Mantovani98}.

Under the condition of  an isotropic distribution of material, and
a continuous precessing jet (period of precession is constant), a
bent jet would precess in space with an unchanged shape, as shown
in the three curves through $A_2$, $B_2$ and $C_2$ respectively in
Fig.~2.

Consequently, the strange morphology of 0735+178 can be explained.
The three hot spots along the bent jet can be treated as three
individual straight jets, with different misalignment angles to
the axis of precession cone, and initial precession phases, as the
dashed lines in Fig.~2. In such case the jet motion of 0735+178
can be fitted as Fig.~6 through the parameters in Table~2.


\begin{table}
\begin{center}
\caption{The fitting parameters of 0735+178
}
\begin{tabular}{cccccc}
\hline \hline
  $\dot{\Omega}$ & $i$ & $\xi$ & $\lambda_1$
  &  $\lambda_2$
&  $\lambda_3$ \\
  19.6 & 12.6 & 50.9 & 9.2 & 15.5  & 6.9 \\
 $\eta_1$ & $\eta_2$ & $\eta_3$ &
 $R_1$ &  $R_2$ &  $R_3$ \\
-154.7 & -166.2 & -17.2
 & 4.0 & 8.4 & 13.1 \\
\hline
\end{tabular}
\end{center}
{\small $\dot{\Omega}$ is in unit of deg/yr. $R_i$ is in unit of
mas, and others are in unit of degree. }
\end{table}

Obviously  the motion of ejecta of 0735+178 is more complicated
than that of 0119+041, because the former contains three hot spots
along a bent jet, whereas the latter contains one.

From the context of the non-ballistic model, there are two main
discrepancies between GRS 1915+105 and 0735+178, firstly the
former satisfies, $\lambda\approx i$, whereas the latter satisfies
$\lambda\neq i$; secondly   $\theta$ can only be observed in a
small range (a few degrees) in the former, whereas it is observed
in  much larger range in the latter.

The first discrepancy explains why the features of  GRS 1915+105
appear ejecting out from the core, whereas that of  0735+178
don't. And the second discrepancy explains  why the trajectories
of GRS 1915+105 appear linear and that of  0735+178 appear
non-linear. In fact a curved trajectory in GRS 1915+105 is
inevitable, if the ejecta of it could be observed at large
$\theta$.

Notice that
the fitting parameters  of 0119+041 and that of Table~1 and
Table~2 for  GRS 1915+105 and 0735+178 respectively  are not
unique. Other numerical solutions to them cannot be excluded.

\begin{figure}
\includegraphics[angle=-90,width=0.45\textwidth]{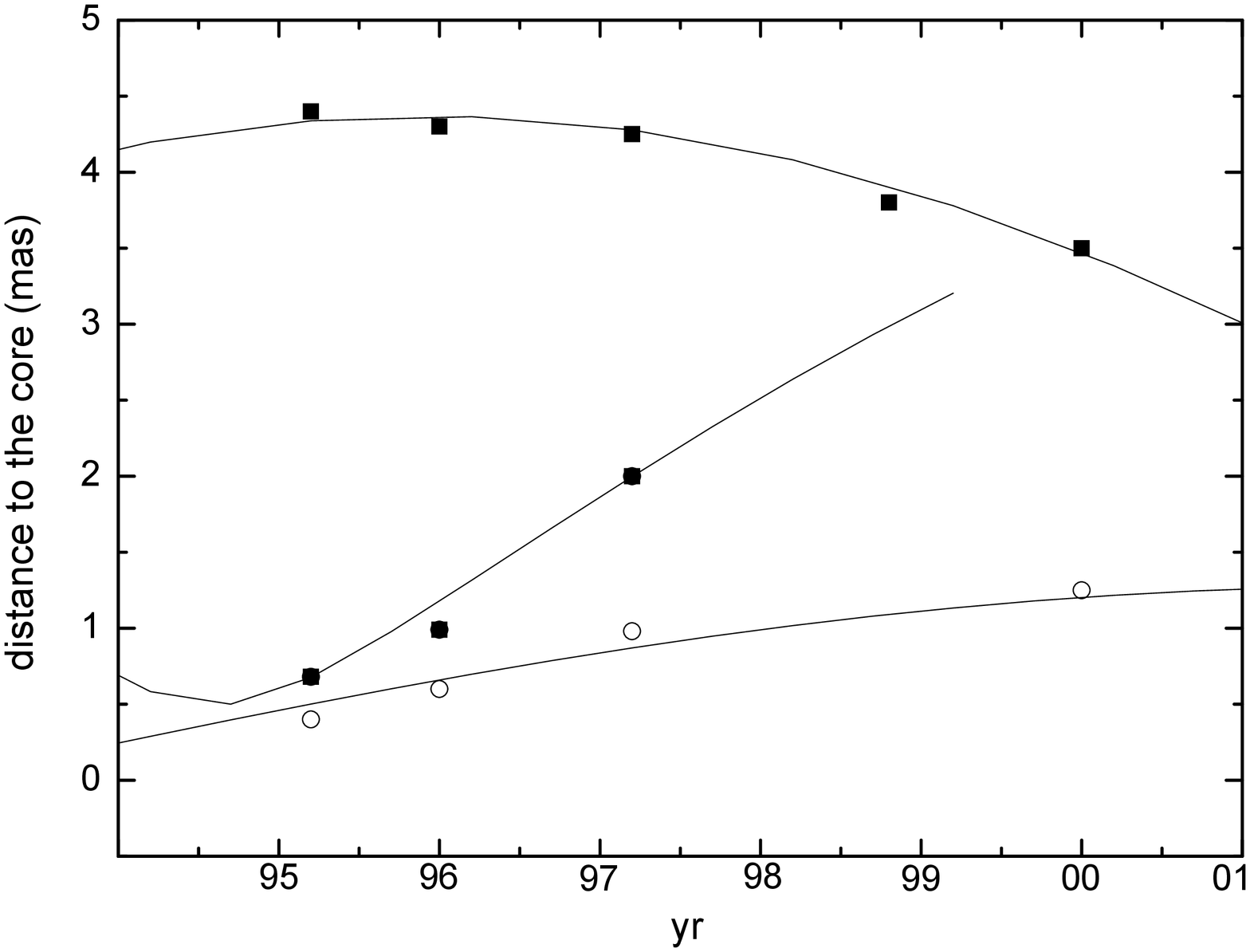}
\caption{\small Fitting of 0735+178. The three hot spots along the
bent jet can be treated as three individual straight jets,
precessing with the same  velocity around the axis of precession
cone, but with different misalignment angles with respect to the
axis of precession cone, and different initial precession phases.
The trajectories of ejecta are fitted by three curves via the
parameters of 0735+178 in Table~2. \label{Fig:fig6} }
\end{figure}

\section{Outward vs inward} \label{sec:outin}

The new model can account for the very large numbers of observed
outward motions, as opposed to the inward ones.

As shown in  Fig.~7,  the  dash profile corresponds to the
apparent luminosity, $L$, obtained by putting $\theta$ of
Eq.~($\ref{theta}$) into Eq.~($\ref{L}$). Since $\theta$ is a
function of time (due to $\eta$), $L$ is dependent of $t$ either.
In fact the profile of $L(t)$, or the light-curve (L-C), can only
be obtained in the case that an emission beam with negligible
opening angle precesses through LOS, in other words, the
convolution of $L(t)$ with a $\delta(t)$ function.

Generally, the L-C appears as  the convolution of  $L(t)$ with the
emission beam, $P(t)$, which has certain opening angle. In such
case, the L-C is given by the convolution,
\begin{equation} \label{conv}
F(t)=\int^{t_b}_{t_a}L(\tau)P(t-\tau)d\tau
 \,,
\end{equation}
where $P(t)$ can be both  homogeneous or nonhomogeneous. Here its
shape  is assumed to be a rapid rise and slow decrease, which can
be denoted by i.e., the triangle at the bottom of Fig.~7. By
$t_a=-\infty$ and $t_b=t_1$, the L-C of an  inward feature is
obtained. Similarly, by setting  $t_a=t_2$ and $t_b=\infty$, the
L-C of the outward feature can be obtained either.

For simplicity, the core emission is assumed to be a constant, as
the horizon line with a hight of unit shown in Fig.~7. However the
jet emission can be attributed to the emission of the core, when
the jet precesses to the small $\theta(t)$ region within the core
area. The core area is between the time interval, $t_1$ to $t_2$,
as shown in the slash rectangular. Thus the apparent luminosity of
the core area becomes, $F(t)+1$, where $F(t)$ is given by setting
$t_a=t_1$ and $t_b=t_2$ in Eq.~($\ref{conv}$).




By Eq.~($\ref{conv}$), the apparent luminosity of the core, the
inward and the outward features are obtained.
The apparent luminosity of an inward feature increases gradually.
It disappears in the core area (region 1), at its maximum $F(t)$,
as shown in the left hand side curve of Fig.~7. The region 1 is
the common area of the core and the inward jet, in which the
decrease of the jet apparent luminosity occurs during the increase
of that of the core, which makes this region difficult to
distinguish from the core.

Contrarily, an outward feature starts in the region 2, which is
beyond the true core area. The jet emission rises rapidly, after
reaching its maximum $F(t)$, it decreases gradually.

Therefore, compare with the inward features, the outward ones have
an additional region 2, which is favorable to observe.  This may
explain the phenomenon of very large numbers of observed outward
motions, as opposed to the inward ones.


The radio L-C of the SE component of GRS 1915+105 at 8.46GHz
(Miller-Jones et al. 2007), as shown in the small box (A, B and C)
is  consistent with the outward curve ($A^{\prime}$, $B^{\prime}$,
and $C^{\prime}$) in Fig.~7.



\begin{figure}
\includegraphics[width=0.43\textwidth]{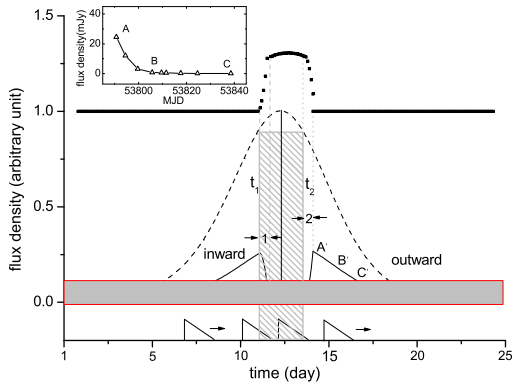}
\caption{\small The  dash profile corresponds to the apparent
luminosity, $L(t)$. The convolution of  $L(t)$ with  the emission
beam, $P(t)$, represented by the triangle at the bottom gives
following results: the apparent luminosity the core, the inward
and the outward features, as shown by the top dashed curve, the
bottom left hand side curve, and the bottom right hand side curve
respectively. The hight of the light gray rectangular at the
bottom represents the limit of telescope sensitivity. The core
area is denoted by the slash rectangular between $t_1$ and $t_2$.
The small box at the top left is the radio L-C of the SE component
of GRS 1915+105 at 8.46GHz.
 \label{Fig:fig7} }
\end{figure}


\section{Discussions and Predictions }

The non-ballistic  model explains superluminal motion by the
apparent transverse motion of hot spots, resulted in  a continuous
jet interacting with material at different locations and
directions at   approximately the same distance to the core.
Obviously, the scenario doesn't mean the transportation of energy
and moment from one point to another with the apparent speed.


The appearance on the sky of  precessing jets is  affected by time
delays, since the image observed is composed of photons coming
from various parts of the jet that happen to have the same arrival
times. However, in the fitting of jet motion of quasars and
microquasars using the non-ballistic model of this paper, the time
delay is neglected. The reason is as follows.

The time delay is determined by the distance along LOS, $R_x$,
giving by the first equation of Eq.~($\ref{lk}$), which is
dependent of angles, $i$, $\eta$ and  $\lambda$, and distance,
$R_d$.

By Dermer (1999) the time corresponding to the radio peak is
around $10^6$ to $10^7$ seconds, which means the distance of the
hot spot to the core is, $R_d=10^{16}$ to $10^{17}$ cm.

Consequently, the discrepancy in the time delay corresponding to
the two hot spots, i.e., $A_2$ and $B_2$ in Fig.~2, can be given
by putting two different precession phases, $\eta(t_1)$ and
$\eta(t_2)$, corresponding to two different time, $t_1$ and $t_2$,
respectively, into the first equation of Eq.~($\ref{lk}$),
\begin{equation} \label{timedelay}
\Delta T=\frac{R_{x}(t_2)-R_{x}(t_1)}{c}
 \,.
\end{equation}
Therefore, by the parameters of  0119+041 and 0735+178 in the
text, the time delay between two measured hot spots is less or
around 1 day respectively, which is negligible in the fitting of
Fig.~3 and Fig.~6. For GRS 1915+105 the time delay is $\sim 3$
hours which doesn't affect the fitting of Fig.~5 either.


If $R_d$  exceeds 100pc, then the time delay  between two measured
hot spots would be larger  than 1 year. Under such circumstance,
the hot spot, i.e., B2 in Fig.~2, would had been emerged at least
1 year before it was observed. In such case, increasing the jet
precession speed, $\dot{\Omega}$, by a few percent, the data can
still be fitted.




The two models can be tested simply by reanalyzing the data of
0119+041. If the previous features $A$ and $B$ move further away
from the core when $C$ is observed, then the ballistic model is
supported, and the features should be produced by discrete jets.
Otherwise, if any intermediate feature is found, i.e., between $A$
and $B$, or $B$ and $C$, then the non-ballistic model is favored,
and the features should be powered by a continuous jet.

The jet motion corresponding to a full period of precession is
expected  on sources like 0119+041 and 0735+178. If this is
observed then the non-ballistic model would be supported.

The discrepancy of the inward and outward feature can also be
tested.  The apparent luminosity of an inward feature increases
gradually, and disappears in the core area (region 1), whereas
that of an outward feature may decrease from a peak.  And the core
area appears moving towards the outward direction for certain
angular distance, as  shown in Fig.~7.

The ballistic model is  based on the assumption of a discrete jet,
and can easily explain the linear jet motion. Comparatively, the
non-ballistic model is based on the continuous jet, which
interprets both the linear and nonlinear jet motion. Therefore, it
explains both the phenomena the ballistic model can explain, and
the ones  the ballistic model cannot.



Testing the  two models through  their different predictions on
both  quasars and microquasars, and  by both the data from the
past and from the future, would enhance our understanding of the
true mechanism of superluminal motion, the interaction of jets
with interstellar matter, and the common physics underlying
quasars and microquasars.


\section{Acknowledgments}
I thank S.N. Zhang, Z.Q. Shen and X. Liu for useful discussions
and suggestions. This research is supported by the National
Natural Science Foundation of China, under grand NSFC10778712.



\end{document}